\documentclass[12pt,letterpaper]{article}

\usepackage[normalem]{ulem}
\usepackage{comment}
\usepackage{multirow}
\usepackage{xpatch}
\makeatletter
\xpatchcmd{\ps@headings}{\sectionmarkformat\fi}{\sectionmarkformat}{}{} 

\usepackage{amsfonts,bm,graphicx,hyperref,xcolor,mathtools}
\usepackage{amsmath,amssymb,ulem}
\usepackage{comment}
\usepackage{amssymb}
\usepackage{enumitem}
\usepackage{mathptmx}
\usepackage{color}
\usepackage[margin=1in]{geometry}
\usepackage{amsmath}
\usepackage{hyperref}
\usepackage{wrapfig}
\usepackage{graphicx}
\usepackage{upgreek}
\usepackage{textcomp}
\usepackage[utf8]{inputenc}
\usepackage[font=small,labelfont=bf]{caption}[2003/12/20]
\usepackage{sectsty}
\usepackage{lineno}
\usepackage[compact]{titlesec}
\usepackage{xcolor}
\usepackage{sidecap}
\usepackage[strict]{chngpage}
\usepackage{braket}
\usepackage[sort&compress,numbers]{natbib}

\usepackage{array}

\newcommand{\til}{{\raise.17ex\hbox{$\scriptstyle\sim$}}}

\newcommand{\pbar}{\ensuremath{\overline{\mathrm{p}}}}
\newcommand{\nbar}{\ensuremath{\overline{\mathrm{n}}}}
\newcommand{\dbar}{\ensuremath{\overline{\mathrm{d}}}}
\newcommand{\hebar}{\ensuremath{\overline{\mathrm{He}}}}

\usepackage{newunicodechar,graphicx}
\DeclareRobustCommand{\okina}{%
  \raisebox{\dimexpr\fontcharht\font`A-\height}{%
    \scalebox{0.8}{`}%
  }%
}
\newunicodechar{ʻ}{\okina}

\makeatletter
\renewenvironment{description}%
               {\list{}{\leftmargin=10pt 
                        \labelwidth\z@ \itemindent-\leftmargin
                        }}%
               {\endlist}
\makeatother

\usepackage{leading}
\leading{14.23pt}

\allsectionsfont{\normalfont\sffamily\bfseries}

\def\be{\begin{equation}}
\def\ee{\end{equation}}
\def\bi{\begin{itemize}[noitemsep,leftmargin=*]}
\def\benum{\begin{enumerate}[noitemsep,leftmargin=*]}
\def\bd{\begin{description}[leftmargin=*]}
\def\ei{\end{itemize}}
\def\bmx{\begin{matrix}}
\def\emx{\end{matrix}}

\titlespacing{\section}{0pt}{5pt}{3pt}
\titlespacing{\subsection}{0pt}{5pt}{3pt}
\titlespacing{\subsubsection}{0pt}{15pt}{3pt}
\titlespacing{\paragraph}{0pt}{10pt}{10pt}
\titlespacing{\subparagraph}{0pt}{10pt}{10pt}

\title{Low-energy antinuclei measurements for background-free indirect dark matter searches and PBH signatures}
\author{\small Tsuguo Aramaki (Northeastern University), Ilias Cholis (Oakland University), \\  \small Philip von Doetinchem (University of Hawaii at Manoa), Fiorenza Donato (University of Torino/INFN), \\ \small Priyarshini Ghosh (University of Maryland, Baltimore County), \\ \small Dan Hooper (University of Wisconsin–Madison), Timothy Linden (Stockholm University), \\ \small Keith McBride (University of Chicago), Stefano Profumo (University of California, Santa Cruz), \\ \small Stephanie Wissel (Pennsylvania State University) \\
\small For PhysPAG\\
}
\date{}

\begin{document}
\clubpenalty = 10000 \widowpenalty = 10000 \displaywidowpenalty = 10000

\maketitle

\section{Science Investigation}

Cosmic ray (CR) antinuclei (here: antiproton (\pbar), antideuteron (\dbar), and antihelium (\hebar)) constitute powerful probes for indirectly detecting candidates for the dark matter (DM) in our Galaxy. The detection of a flux of antiprotons between 1 and 400~GV, performed with great accuracy by AMS-02 \cite{PhysRevLett.117.091103,AMS:2016oqu,AMS:2021nhj}, demonstrates that nuclear antimatter is produced in the Galaxy. While these precise data across intermediate and high rigidities have been shown to be in statistical agreement with a purely secondary origin due to the fragmentation of incoming CR nuclei over the atoms of the interstellar medium (ISM)
 \cite{Boudaud:2019efq, Heisig:2020nse, Balan:2023lwg, DiMauro:2023jgg, Calore:2022stf, DelaTorreLuque:2024ozf, Stefanuto:2026wxy},
a small contribution due to exotic primary sources cannot be excluded \cite{Heisig:2020nse,Balan:2023lwg,Stefanuto:2026wxy}. 

The secondary production of antinuclei 
proceeds via high-energy fixed-target collisions in the Galactic disk
such as $p + p \to p + p + p + \bar{p}$ and $p + p \to p + p + \bar{d} + X$, etc. 
To conserve baryon number, these interactions demand substantial threshold kinetic energies in the laboratory frame: $K_{\text{th}} \approx 5.6~\text{GeV}$ for \pbar\ and $K_{\text{th}} \approx 15~\text{GeV}$ for \dbar\ (see Table I in \cite{Stefanuto:2026wxy}). 
Furthermore, because the center-of-mass frame moves at highly relativistic velocities relative to the Galactic rest frame, the generated antimatter products inherit significant forward momentum. This kinematically shifts the peak of the secondary background spectrum toward $2\text{--}5~\text{GeV}$ for \pbar\ and $4\text{--}7~\text{GeV}/n$ for \dbar. Below $1~\text{GeV}/n$, the secondary background flux drops precipitously for antinuclei. 

\subsubsection*{Primary Production in the Galaxy:  Particle Dark Matter and Primordial Black Holes }

In contrast to secondary production, primary cold DM relics (see \cite{Cirelli:2024ssz} for a recent review) —such as Weakly Interacting Massive Particles (WIMPs)—annihilate or decay within the Galactic halo with non-relativistic velocities ($v \sim 10^{-3}c$). The center-of-mass frame of these interactions is effectively identical to the Galactic rest frame, allowing a significant fraction of the hadronization products to be injected at low kinetic energies. 
This low-energy window establishes an exceptionally high signal-to-noise environment for primary sources whose parent particles interact or decay essentially at rest, looking for \dbar\ \cite{Donato:1999gy, Baer:2005tw,Donato:2008yx,Stefanuto:2026wxy} and \hebar\ \cite{Carlson:2014ssa, Cirelli:2014qia}. 

The formation of composite antinuclei from the underlying partonic cascade is modeled using coalescence frameworks. 

Modern data-driven treatments often implement advanced quantum mechanical multi-body Wigner formalisms, reducing the underlying nuclear physics uncertainties that historically plagued low-energy fluxes \cite{Kachelriess:2020uoh,DiMauro:2025vxp,Stefanuto:2026wxy}. 
These predictions, however, have been tested with data at LHC energies. Data for fusion at the CR energies at stake are expected soon by the NA61/SHINE and AMBER Collaborations \cite{Maurin:2025gsz}. 

Primordial black holes (PBHs), assumed to have formed via highly overdense fluctuations in the early universe, present a completely independent, non-thermal channel for producing low-energy cosmic-ray antimatter. According to Hawking's semi-classical mechanism, a black hole continuously radiates particles across all available quantum degrees of freedom, behaving as a blackbody with an intrinsic temperature $T_{\text{BH}}$ inversely proportional to its mass $M_{\text{BH}}$ \cite{Barrau:2001ev, Ukwatta:2015iba}.

As a PBH approaches the end of its lifetime 
($\text dM_{\text{BH}}/\text d t \propto - M_{\text{BH}}^{-2}$), its temperature climbs well past the Quantum Chromodynamics (QCD) confinement scale ($\Lambda_{\text{QCD}} \approx 200\text{--}300~\text{ MeV}$). At this stage, the PBH directly emits fundamental quarks and gluons via explosive jet production \cite{Barrau:2001ev, DeRomeri:2025dwm}. These partons rapidly fragment into stable hadrons, yielding a primary injection of \pbar\ and \nbar,\ which undergo coalescence, producing a flat, distinctive low-energy \dbar\ and \hebar\ spectrum in the sub-GeV domain that uniquely maps onto the local PBH evaporation density 
\cite{Auffinger:2022khh, DeRomeri:2025dwm}.

DM signatures are intrinsically tied to the particle mass scale $m_{\chi}$, producing smooth, characteristic bumps that vary according to the primary hadronization channel (e.g., $b\bar{b}$ or $W^+W^-$). Conversely, PBH evaporation generates a non-thermal spectrum integrated over a mass distribution, which can yield flatter or uniquely sloped low-energy features dictated by Hawking temperature dynamics \cite{DeRomeri:2025}. Critically, any primary flux predicted for heavier antinuclei like \dbar\ or \hebar\ must abide by the stringent constraints imposed by the exceptionally precise AMS-02 \pbar\ data \cite{AMS:2021nhj}. 
Models attempting to explain tentative low-energy \dbar\ or $\overline{\text{He}}$ excesses are heavily bounded by the risk of overproducing antiprotons \cite{DelaTorreLuque:2023,Stefanuto:2026wxy}. 

The negligible astrophysical background for antinuclei below a few GeV/$n$ provides a unique experimental probe of dark matter (DM) models constrained by  \pbar\ data, while also offering the potential to reveal DM signatures that would remain inaccessible through antiproton observations alone. For example, in \cite{DiMauro:2026owr} it has been shown that if DM annihilates into a confining dark sector that produces soft unclustered energy patterns, DM annihilation can yield few or even tens of events already at GAPS or AMS-02.  
A future confirmation of an antinuclei signal by low-energy
experiments could provide strong hints for hidden confining dynamics. 
Establishing a robust exotic origin requires global multi-species fits that simultaneously explain \pbar\, \dbar\, and \hebar\ observables, combined with external multi-messenger bounds from cosmic microwave background anisotropies and diffuse gamma-ray backgrounds.

\subsubsection*{Multi-messenger signals:  gamma rays and antinuclei}
An excess of GeV-energy gamma-rays toward the center of our Galaxy, known as the Galactic Center Excess (GCE), has been robustly observed by the Fermi Large Area Telescope~\cite{Goodenough:2009gk, Vitale:2009hr, Hooper:2010mq, Abazajian:2010zy, Gordon:2013vta, Calore:2014xka, Zhou:2014lva, Fermi-LAT:2015sau, DiMauro:2021raz, Cholis:2021rpp}, and, more recently, independently confirmed by DAMPE telescope data \cite{DAMPE:2025ndf}. 
The three basic explanations proposed for its origin are
that (i) it is a signal of an unresolved population of millisecond pulsars (MSPs) at the Galactic Center~\cite{Abazajian:2012pn, Petrovic:2014xra, Lee:2015fea, Bartels:2017vsx, Gautam:2021wqn, Macias:2023qqc, Calore:2021jvg,Manconi:2024tgh,Manconi:2024tgh}, (ii) it is produced primarily from DM annihilations \cite{Hooper:2011ti, Hooper:2013rwa, Gordon:2013vta, Daylan:2014rsa, Calore:2014xka, Calore:2014nla, Agrawal:2014oha, Berlin:2015wwa, Fermi-LAT:2017opo, Karwin:2016tsw, Leane:2019xiy, DiMauro:2021raz, Cholis:2021rpp, Zhong:2024vyi}, and (iii) it is the result of bursts of CRs possibly associated with the supermassive black hole's environment \cite{Petrovic:2014uda, Carlson:2014cwa, Cholis:2015dea}.

However, observations reveal some tension with each scenario: 
the $\gamma$-ray luminosity function of MSPs is not well defined, 
as well as our best understanding of their expected radio and X-ray signals~\cite{Hooper:2013nhl, Cholis:2014noa, Cholis:2014lta, Zhong:2019ycb, Buschmann:2020adf, Hooper:2021kyp, Dinsmore:2021nip, List:2025qbx}. 
In scenarios where the GCE is a signal from WIMP DM, the associated signals would be expected in other targets such as $\gamma$-ray searches in dwarf spheroidal galaxies, Andromeda and the isotropic $\gamma$-ray background, as well as in cosmic-ray studies~\cite{Hooper:2014ysa, Cirelli:2014lwa, Bringmann:2014lpa,Calore:2022stf}. While no such signal has been definitively detected (leading to some tension in the case of dwarf galaxy observations~\cite{McDaniel:2023bju}), a mild signal was reported in 
the antiproton flux \cite{Cuoco:2016eej, Cui:2016ppb, Cholis:2019ejx}, even if the inclusion of correlations in the error matrix blows out the hint \cite{Boudaud:2019efq, Heisig:2020nse, Balan:2023lwg, Calore:2022stf, DelaTorreLuque:2024ozf,Stefanuto:2026wxy}.

The correlation between gamma-ray signals and antiprotons from DM annihilation can be very strong, in particular when the DM particles annihilates into light quarks. 
It was shown that in case the GCE is due to a WIMP with 
$m_\chi \simeq$ 50--90 GeV, a contribution at the $O(0.1)$ level would be present in CR \pbar\  flux at 5-20 GeV/$n$ \cite{Cholis:2019ejx, Cuoco:2019kuu}. 
Furthermore, a WIMP of that same mass range, would provide a significant higher $^3$\hebar\ with respect to secondary flux at low kinetic energies, allowing for the detection of these otherwise unobservable CR species \cite{Kachelriess:2020uoh, Cholis:2020twh,Winkler:2020ltd,Stefanuto:2026wxy}. 
With a good understanding of future detector's sensitivity to \dbar\ and \hebar\, and the local CR propagation conditions, given how suppressed the secondary fluxes are, the measurement of the ratio of $^3$\hebar\ to \dbar\ CR  fluxes can be used to test and potentially discover DM  candidates \cite{Kachelriess:2020uoh, Cholis:2020twh, Winkler:2020ltd, Winkler:2022zdu, Ding:2022bgp, DeLaTorreLuque:2024htu,Stefanuto:2026wxy}.

\section{Science Table}

\begin{table*}[htbp]
\centering
\footnotesize
\renewcommand{\arraystretch}{1.3}
\begin{tabular}{|>{\raggedright\arraybackslash}p{0.21\textwidth}|
                >{\raggedright\arraybackslash}p{0.24\textwidth}|
                >{\raggedright\arraybackslash}p{0.24\textwidth}|
                >{\raggedright\arraybackslash}p{0.22\textwidth}|}
\hline
\textbf{Physical Parameters} &
\textbf{Science Objectives} &
\textbf{Observables} &
\textbf{Potential Challenges} \\
\hline
Primary antideuteron detection
&
Potential DM and PBH discovery
&
Sensitivity at $<~10^{-5}$~[m$^2$sr(GeV/$n$)s)]$^{-1}$
&
Need to reject antiproton events
\\
\hline
Secondary antideuteron detection
&
Provide constraints on antideuteron production models (cross section and coalescence model) and DM/PBH contribution
&
Sensitivity at $<~10^{-8}$~[m$^2$sr(GeV/$n$)s)]$^{-1}$
&
Large acceptance area and long measurement times are required. 
\\
\hline
Primary antihelium detection
&
New, unknown sources; Potential DM and PBH discovery
&
New, unknown sources: Sensitivity at $<~10^{-9}$~[m$^2$sr(GeV/$n$)s)]$^{-1}$
Potential DM and PBH discovery: Sensitivity at $<~10^{-11}$~[m$^2$sr(GeV/$n$)s)]$^{-1}$
&
Large acceptance area and long measurement times are required. 
\\
\hline
Secondary antihelium detection
&
Provide constraints on antihelium production models (cross section and coalescence model) and DM/PBH contribution
&
Sensitivity at $<~10^{-12}$~[m$^2$sr(GeV/$n$)s)]$^{-1}$
&
Large acceptance area and long measurement times are required. 
\\
\hline
\end{tabular}
\end{table*}

\section{Instrument Description\label{s-inst}}

\begin{figure}
    \centering
    \includegraphics[width=0.9\textwidth]{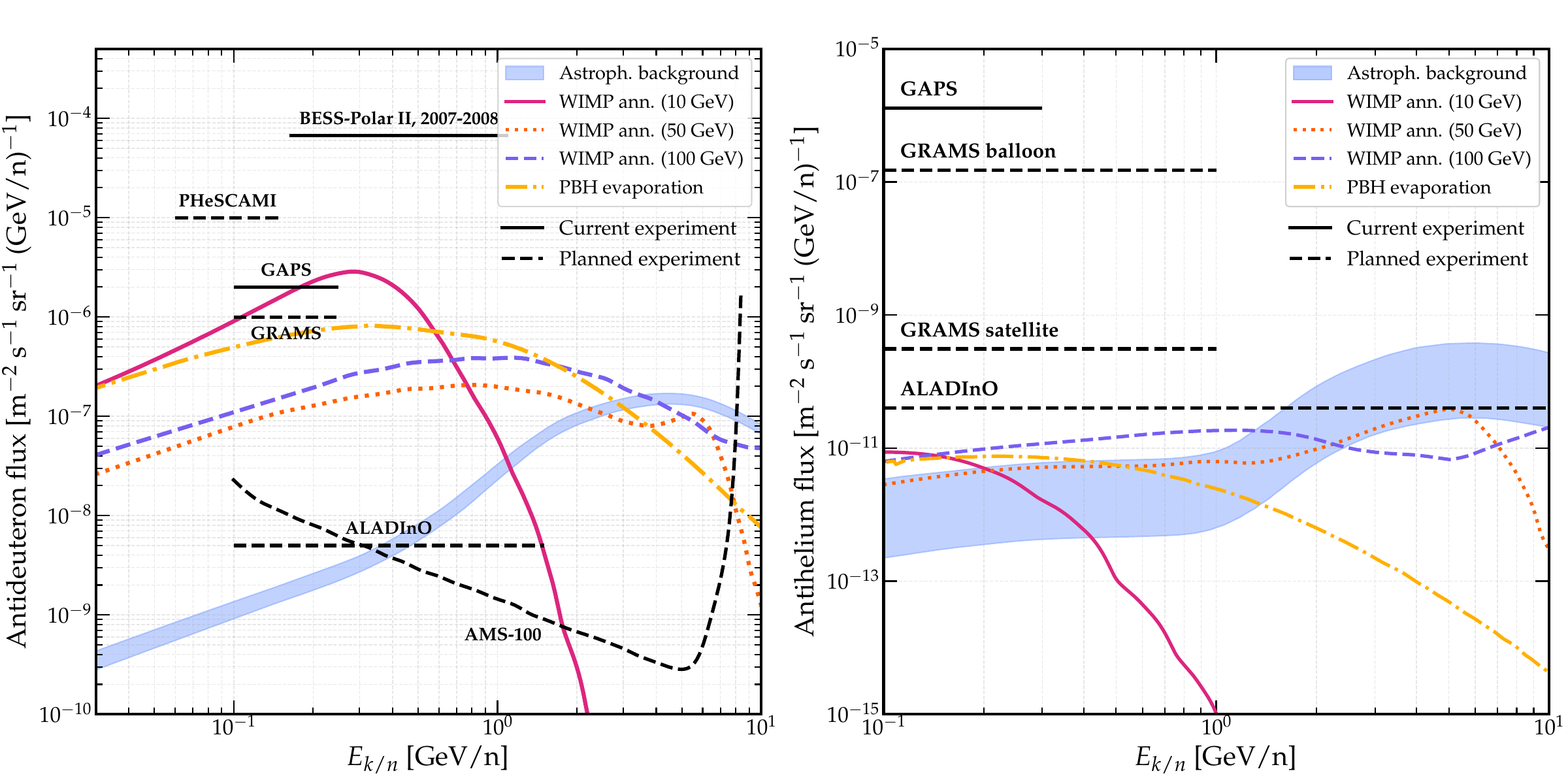}
    \vspace{-0.15in}
    \caption{Antideuteron (left) and antihelium nuclei (right) sensitivities for current and next-generation missions. 
Examples for DM or PBH models: Solid red/dotted orange /dashed blue lines indicate the flux from $m_\chi=10 / 50 /100)$ GeV DM annihilation, while the dot-dashed line is for PBH evaporation. AMS-02 antiproton data constrain all the models. 
    The secondary flux, with its uncertainty band, is depicted by the blue band. 
    All the predictions are from \cite{Stefanuto:2026wxy} except for the $^3$\hebar\ flux, which is taken from PBH and is from \cite{Herms:2016vop}. 
    }
    \label{fig:Sensitivities}
    \vspace{-0.2in}
\end{figure}

The detection concept and particle identification techniques for current and next-generation antinuclei measurement missions are based either on the creation and detection of exotic atoms, in which antinuclei replace shell electrons, or on measuring antinuclei trajectories in strong magnetic fields. The sensitivities to antideuterons and antihelium nuclei for the current and proposed missions are shown in Fig.~\ref{fig:Sensitivities}.
Concerning antihelium modeling, it has to be noted that the reported antihelium candidates by AMS-02~\cite{2022cosp...44.2083C} can currently not be explained within DM or PBH models, and must be orders of magnitude above the prediction for the astrophysical background.

\subsubsection*{Particle identification with exotic atom decay}

The antinuclei identification method, based on the decay of the exotic atom and its annihilation products, has been developed for the GAPS project\cite{Mori_2002,Aramaki_2013,Aramaki_2016}. The GAPS Antarctic long-duration balloon experiment \cite{thegapscollaboration2026generalantiparticlespectrometergaps} focuses on the detection of low-energy antinuclei by stopping antinuclei in its silicon tracker, and subsequent creation of an exotic atom. By tracking the decay and annihilation products of the unstable exotic atom, antinuclei species can be identified.
GAPS is the current-generation experiment and completed its first flight from the Antarctic in January 2026, and is currently analyzing the data. Two more flights are planned to be completed within the next 10 years. Until the mid-2030s, GAPS will be the only dedicated low-energy cosmic-ray experiment with exposure to cosmic antinuclei. 
The GAPS detection concept could also be an attractive add-on for next-generation multipurpose cosmic-ray lunar or L2 platforms.

GRAMS (Gamma-Ray and AntiMatter Survey) \cite{Aramaki_2020} is a next-generation balloon-borne and satellite-based mission that uses the same detection technique as GAPS but employs a liquid argon time projection chamber (LArTPC) detector instead of a silicon tracker. In addition, the GRAMS detector functions as a Compton camera to measure gamma rays in the important MeV energy band. GRAMS conducted an engineering flight from Taiki, Japan, in 2023 \cite{Nakajima_2024} and an antiproton beam test in 2025 \cite{Yano_2026}. The GRAMS collaboration is currently preparing for a prototype balloon flight from Arizona in the Fall of 2026 to demonstrate the detector's performance in flight. The first Antarctic science flight is envisioned in the next decade, subject to funding and mission approval. As a potential future outlook, a GRAMS satellite concept at L2 is also discussed \cite{Zeng_2025}.

PHeSCAMI (Pressurized Helium Scintillating Calorimeter for AntiMatter Identification, also known as ADHD (Anti-Deuteron Helium Detector) is another proposed mission utilizing a high-pressure helium calorimeter. The incoming particle forms a metastable exotic atom with the helium nucleus, where the delayed signal from the decay of the metastable exotic atom and the number of annihilation products will be used to identify the incoming antinuclei.

\subsubsection*{Particle identification with particle trajectory in a strong magnetic field}

Space-based and balloon-borne experiments, like BESS, PAMELA, and AMS-02 \cite{Fuke:2005it,Sakai_2024,2012PhRvL.108m1301A,besspbar, Adriani:2012paa, PAMELA:2017bna, PhysRevLett.117.091103, PhysRevLett.132.131001} have been searching for cosmic antinuclei for decades.
They measure the charge, rigidity, and mass by analyzing particle trajectories in a strong magnetic field.
AMS-02 is the current-generation mission and has been operating on the International Space Station since 2011 and will be operational until the end of the ISS, which is currently scheduled for 2030. It is a multi-purpose cosmic-ray detector that includes a permanent magnet and a silicon tracker. The experiment precisely measured low-energy antiprotons down to sub-GeV while also reporting the detection of antihelium candidates in the energy region of a few GeV/$n$ with a rate of about one per year, and a similar amount of antideuteron candidates in the same energy range~\cite{2022cosp...44.2082C, 2022cosp...44.2083C}. 

ALADInO (Antimatter Large Acceptance Detector In Orbit) \cite{Battiston:2021org} and AMS-100 (A Magnetic Spectrometer with a geometrical acceptance of 100\,m$^2$sr) \cite{2019NIMPA.94462561S} are next-generation space-based mission concepts with significantly increased acceptance for cosmic rays, including low-energy antinuclei, up to the PeV energy scale, compared to AMS-02. They consist of various subdetectors, including tracking and time-of-flight systems, for particle identification of bent tracks in a strong magnetic field generated by a high-temperature superconducting (HTS) magnet. They are designed to operate at the Sun-Earth Lagrange Point 2 (L2) to minimize the geomagnetic cutoff effect and benefit from an ideal environment for the HTS magnet, with an estimated launch of around 2040.

\section{Mission Implementation}

To guarantee exposure to low-energy antinuclei, the preferred mission destinations are Antarctic balloon flights for current and mid-term missions (maximum of about 100\,days). However, to push the sensitivity to low-energy antinuclei by orders of magnitude, satellite missions at the Earth–Sun Lagrange Point L2 (or in high Earth orbit) for about 5--10 years are required.
L2 also has the advantage of avoiding atmospheric attenuation effects and provides an effective environment for HTS magnets and cryogenic detectors. Below is the summary table for the current and next-generation missions.

\vspace{-0.1in}
\begin{table*}[htbp]
\centering
\footnotesize
\renewcommand{\arraystretch}{1.0}
\begin{tabular}{|> 
{\centering\arraybackslash}p{0.12\textwidth}|
                >{\centering\arraybackslash}p{0.18\textwidth}|
                >                {\centering\arraybackslash}p{0.24\textwidth}|
                >                {\centering\arraybackslash}p{0.22\textwidth}|
                >{\centering\arraybackslash}p{0.08\textwidth}|
                }
\multicolumn{5}{c@{\hspace{12cm}}}{Current-generation} \\
\hline
\textbf{Mission} &
\textbf{Detection concept} &
\textbf{Detector technology} &
\textbf{Mission destination} &
\textbf{Ref} \\
\hline
\hline
BESS & Particle trajectory & Superconducting magnet & Balloon (Antarctic) &\cite{Fuke:2005it,Sakai_2024}\\
\hline
AMS-02 & Particle trajectory & Permanent magnet & Satellite (ISS) &\cite{2022cosp...44.2082C, 2022cosp...44.2083C}\\
\hline
GAPS & Exotic atom decay & Silicon tracker & Balloon (Antarctic) &\cite{thegapscollaboration2026generalantiparticlespectrometergaps}\\
\hline
\multicolumn{5}{c@{\hspace{12cm}}}{} \\

\multicolumn{5}{c@{\hspace{12cm}}}{Next-generation} \\
\hline
\textbf{Mission} &
\textbf{Detection concept} &
\textbf{Detector technology} &
\textbf{Mission destination} &
\textbf{Ref} \\
\hline
\hline
GRAMS & Exotic atom decay & LArTPC & Balloon, Satellite (L2) &\cite{Aramaki_2020} \\
\hline
PHeSCAMI & Exotic atom decay & high-pressure helium gas & Satellite &\cite{Nozzoli_2020}\\
\hline
AMS-100 & Particle trajectory & HTS magnet & Satellite (L2) &\cite{2019NIMPA.94462561S}\\
\hline
ALADInO & Particle trajectory & HTS magnet & Satellite (L2) &\cite{Battiston:2021org}\\
\hline
\end{tabular}
\end{table*}

\bibliographystyle{naturemag_noURL}
\bibliography{main.bib}

\end{document}